\documentclass[english,preprint]{iopart}
\usepackage[T1]{fontenc}
\usepackage[latin9]{inputenc}
\usepackage{color}
\usepackage{units}
\usepackage{textcomp}
\usepackage{amstext}
\usepackage{amssymb}
\usepackage{graphicx}
\usepackage{subscript}

\makeatletter

\providecommand{\tabularnewline}{\\}

\usepackage{iopams}
\usepackage{setstack}

\makeatother

\usepackage{babel}
\begin{document}
\begin{flushright}
Version: \today
\par\end{flushright}

\title{Epitaxial (111) Films of Cu, Ni, and Cu\textsubscript{x}Ni\textsubscript{y}
on $\alpha-$Al\textsubscript{2}O\textsubscript{3}(0001) for Graphene
Growth by Chemical Vapor Deposition}

\author{\author{David L. Miller, Mark W. Keller, Justin M. Shaw}
\address{Magnetics Group, Electromagnetics Division}
\author{Ann N. Chiaramonti, Robert R. Keller}
\address{Nanoscale Reliability Group, Materials Reliability Division}}

\author{\address{National Institute of Standards and Technology, Boulder, CO 80305}}

\ead{david.miller@nist.gov, mark.keller@nist.gov}
\begin{abstract}
Films of (111)-textured Cu, Ni, and Cu\textsubscript{x}Ni\textsubscript{y}
were evaluated as substrates for chemical vapor deposition of graphene.
A metal thickness of $\unit[400]{nm}$ to $\unit[700]{nm}$ was sputtered
onto a substrate of $\alpha-$Al\textsubscript{2}O\textsubscript{3}(0001)
at temperatures of $\unit[250]{^{\circ}C}$ to $\unit[650]{^{\circ}C}$.
The films were then annealed at $\unit[1000]{^{\circ}C}$ in a tube
furnace. X-ray and electron backscatter diffraction measurements showed
all films have (111) texture but have grains with in-plane orientations
differing by $60^{\circ}$. The in-plane epitaxial relationship for
all films was\textcolor{black}{{} $[110]_{\textrm{metal}}$||$[10\bar{1}0]_{\textrm{Al}{}_{2}\textrm{O}_{3}}$}.
Reactive sputtering of Al in O\textsubscript{2} before metal deposition
resulted in a single in-plane orientation over 97~\% of the Ni film
but had no significant effect on the Cu grain structure. Transmission
electron microscopy showed a clean Ni/Al\textsubscript{2}O\textsubscript{3}
interface, confirmed the epitaxial relationship, and showed that formation
of the $60^{\circ}$ twin grains was associated with features on the
Al\textsubscript{2}O\textsubscript{3} surface. Increasing total
pressure and Cu vapor pressure during annealing decreased the roughness
of Cu and and Cu\textsubscript{x}Ni\textsubscript{y} films. Graphene
grown on the Ni(111) films was more uniform than that grown on polycrystalline
Ni/SiO\textsubscript{2} films, but still showed thickness variations
on a much smaller length scale than the distance between grains.
\end{abstract}

\address{Official contribution of the National Institute of Standards and
Technology; not subject to copyright in the United States}

\maketitle

\section{Introduction}

Uniform growth of graphene over wafer-scale areas is a critical enabling
step for the commercial realization of various electronic, photonic,
mechanical, and other devices based upon the superlative properties
of graphene \cite{Avouris:2010bh}. Considerable progress has been
made in the growth of monolayer and few-layer graphene by chemical
vapor deposition (CVD) on transition metals, particularly Cu and Ni
\cite{Batzill:2012fk,Mattevi:2011fv}. Typically, the metal surface
is heated to $\unit[1000]{^{\circ}C}$ and exposed to a hydrocarbon
gas such as methane that decomposes catalytically to provide a source
of C atoms on the surface. In the case of Cu, growth occurs almost
entirely on the surface and is limited to a single layer for a wide
range of hydrocarbon partial pressure. In the case of Ni, growth involves
C dissolution into the film at high temperature and then precipitation
to the surface as the film cools, which can result in both monolayer
and multilayer graphene \cite{Reina:2009sd}. Alloys of Cu and Ni
offer control over graphene thickness by adjusting the C solubility
in the film \cite{Liu:2011oq}. The Cu or Ni can be etched away to
transfer the graphene layer to an insulating substrate for further
device fabrication steps \cite{Suk:2011vn}.

Commercially available polycrystalline foils of Cu and Ni are commonly
used for CVD growth of graphene because they can be annealed to have
millimeter grain sizes. The best CVD graphene films to date, in terms
of graphene grain size and carrier mobility, have been achieved on
foils \cite{Li:2010bh}, but single crystal films offer several advantages.
First, the hexagonal symmetry and small lattice mismatch of the (111)
surface should provide a more ideal template for graphene growth than
the (100), (110), or higher index surfaces. This idea is supported
by recent experiments showing less rotational disorder in graphene
grown on Cu(111) than on Cu(100) \cite{Wofford:2010fk,Nie:2011ys}.
Second, films typically provide a much smoother growth surface because
foils have roughness due to the rolling process that is not fully
removed by annealing. The experiments in \cite{Nie:2011ys} show how
surface defects on Cu(111) can cause rotational disorder either at
nucleation or during subsequent growth of graphene islands. Third,
the diffusion of C into and out of Ni is different at grain boundaries
than in the interior of grains, so better growth control and better
uniformity are expected if grain boundaries are eliminated \cite{Zhang:2010qf,Yoshii:2011fk}.
Lastly, films are supported by a flat, rigid substrate, helping to
simplify the process of graphene transfer to other materials. Thus
we anticipate that high quality Cu(111) and Ni(111) films will enable
CVD growth of graphene having improved properties compared to graphene
grown on polycrystalline foils.

A suitable substrate for the metal films must, at a minimum, promote
epitaxial growth with (111) texture and be physically and chemically
stable under graphene CVD conditions. For wafer-scale production of
graphene, the substrate should be commercially available in wafer
form at a reasonable cost. The ability to reuse substrate wafers after
metal etching to release the graphene layer is also desirable. Currently,
the material that appears to best satisfy these requirements is $\alpha-$Al\textsubscript{2}O\textsubscript{3}(0001),
which is widely used by manufacturers of radio-frequency electronics
and light-emitting diodes in the form of wafers with diameters up
to $\unit[200]{mm}$. Given the advantages of single crystal Cu(111)
and Ni(111) films described above, achieving such films on Al\textsubscript{2}O\textsubscript{3}
is an important step toward commercial graphene devices.

Previous investigations of Cu deposited on $\alpha-$Al\textsubscript{2}O\textsubscript{3}(0001),
mainly in ultra-high vacuum environments, found a variety of growth
behaviors depending on how the Al\textsubscript{2}O\textsubscript{3}
surface was prepared. Kelber \emph{et al.} \cite{Kelber:2000fk} summarized
several earlier results, discussed them in terms of multiple Al\textsubscript{2}O\textsubscript{3}(0001)
surface terminations and multiple Cu ionization states, and emphasized
the role of a hydroxyl layer bonded to the Al\textsubscript{2}O\textsubscript{3}
surface. The last point is particularly relevant for the work presented
here. When Al\textsubscript{2}O\textsubscript{3} is exposed to ambient
air, water decomposes to form a hydroxyl (OH) layer \cite{Eng:2000nx}
that is difficult to remove. While bombardment with 200 eV to 1000
eV Ar ions can remove all but $\sim\unit[0.1]{monolayer}$ of OH,
annealing in O\textsubscript{2} at temperatures $\geq\unit[800]{^{\circ}C}$
is required to recrystallize the damaged surface \cite{Oh:2006zr,Scheu:2006bh}.
In typical film deposition chambers where such treatments are not
available, an OH layer is likely present on Al\textsubscript{2}O\textsubscript{3}
substrates and may interfere with epitaxial growth. This may be why
previous Cu films deposited on Al\textsubscript{2}O\textsubscript{3}(0001)
\cite{Reddy:2011kx,Ishihara:2011kx,Hu:2012vn} were not single crystals
but consisted of (111) grains having in-plane orientations differing
by $60^{\circ}$. While Ni on $\alpha-$Al\textsubscript{2}O\textsubscript{3}(0001)
has not been the subject of extensive experimental studies, the OH
layer can also be expected to interfere with epitaxy in this case.
As with Cu, previous Ni films on Al\textsubscript{2}O\textsubscript{3}(0001)
showed in-plane orientations differing by $60^{\circ}$ \cite{Yoshii:2011fk}.

Epitaxy and film adhesion are relevant to graphene growth for several
reasons. For Cu, which has a bulk melting point of $\unit[1084]{^{\circ}C}$,
evaporation can cause roughness or even complete loss of the Cu film
\cite{Ismach:2010bh}. Previous work on Cu \cite{Hu:2012vn} showed
that epitaxial and highly textured films with large grains survive
graphene CVD conditions better than polycrystalline films with small
grains. Ni evaporation under growth conditions is negligible, since
the bulk melting point is $\unit[1453]{^{\circ}C}$, but Ni films
can be damaged by the annealing in H\textsubscript{2} which typically
precedes graphene growth \cite{Thiele:2010rb}. An additional consideration
for Ni is that films thinner than $\unit[100]{nm}$ are desirable
in order to limit the total amount of C that is available to precipitate
to the surface, since $\approx\unit[35]{nm}$ of Ni at $\unit[1000]{^{\circ}C}$
can absorb enough C to form a monolayer \cite{Dunn:1968kx}. Such
thin Ni films, even when epitaxial and highly textured, can be damaged
by graphene CVD conditions \cite{Yoshii:2011fk}. Given the various
factors summarized here, the properties and limitations of Cu and
Ni films must be considered, alongside the kinetics and thermodynamics
of graphene formation from a hydrocarbon precursor, when optimizing
a recipe for graphene growth by CVD.

This paper reports a study of sputtered Cu, Ni, and Cu/Ni films that
have been exposed to typical graphene CVD conditions. We used x-ray
diffraction (XRD) and electron backscatter diffraction (EBSD) to measure
crystallinity, atomic force microscopy (AFM) to measure surface morphology,
transmission electron microscopy (TEM) to image the metal/substrate
interface, and optical microscopy to show film properties over large
areas. We also report results for graphene growth by CVD on these
films, characterized by scanning electron microscopy (SEM) and Raman
spectroscopy.

\section{Film Preparation}

Wafers of $\alpha$-Al\textsubscript{2}O\textsubscript{3}(0001)
were prepared by annealing at $\unit[1000]{^{\circ}C}$ in O\textsubscript{2}
at atmospheric pressure for 24 h to remove scratches due to polishing
and give atomically flat terraces. Chips of $\unit[5]{mm}\textrm{ x }\unit[6]{mm}$
were cleaned by ultrasonic agitation in acetone and isopropanol, mounted
on a resistively heated Cu puck, and placed in a cryopumped vacuum
system with a base pressure below $\unit[10^{-5}]{Pa}$ $(\unit[10^{-7}]{Torr})$.
Metal films were deposited by dc magnetron sputtering from 76 mm targets
of 99.99+~\% Cu or Ni in $\unit[0.67]{Pa}$ $(\unit[5]{mTorr})$
of Ar. Sputtering powers between $\unit[50]{W}$ and $\unit[200]{W}$,
corresponding to deposition rates calibrated using a profilometer
o\textcolor{black}{f $\unit[0.3]{nm/s}$ to $\unit[1]{nm/s}$, pr}oduced
no significant differences in film properties. For some films, a ``seed
layer'' of Al\textsubscript{2}O\textsubscript{3} was deposited immediately
before the metal by reactive sputtering from a 76 mm target of 99.999~\%
Al in a mixture of 40~\% O\textsubscript{2} in Ar (both 99.999~\%)
at a total pressure of $\unit[0.67]{Pa}$ $(\unit[5]{mTorr})$ and
a power of $\unit[50]{W}$. The target-to-substrate distance in all
cases was 10 cm.

The use of a reactively sputtered seed layer to improve film adhesion
and epitaxy was motivated by a desire to remove OH without using ion
bombardment and high temperature annealling. Sputtering with high
O\textsubscript{2 }content causes ``resputtering'' of the substrate
due to bombardment by O\textsubscript{}\textsuperscript{\textendash{}}
ions generated at the target and accelerated across the plasma sheath
\cite{Kester:1993fk}. This results in a low net deposition rate,
$<\unit[0.01]{nm/s}$ in our case, and should remove the OH layer
from the substrate. The TEM results discussed below show that the
reactive sputtering did not damage the surface. For the films reported
here, this step was done at a substrate temperature of $\unit[650]{^{\circ}C}$
to promote crystallinity, but the technique was also successful at
$\unit[500]{^{\circ}C}$.

Metal film properties were strongly influenced by the substrate temperature
during sputtering. The deposition temperature $T_{\textrm{d}}$ reported
here was measured using a thermocouple clamped to the side of the
puck facing away from the sputter gun. For fixed $T_{\textrm{d}}$
and without a seed layer, the films did not adhere to the Al\textsubscript{2}O\textsubscript{3}
substrate above $\approx\unit[400]{^{\circ}C}$ for Cu and $\approx\unit[600]{^{\circ}C}$
for Ni. Most films reported here were deposited at $T_{\textrm{d}}=\unit[275]{^{\circ}C}$
for Cu and $T_{\textrm{d}}=\unit[475]{^{\circ}C}$ for Ni. Cu/Ni bilayer
films were formed by depositing $\unit[50]{nm}$ of Ni and then $\unit[350]{nm}$
of Cu. These films formed a homogeneous alloy upon annealing (see
below) and we refer to them as ``Cu-Ni'' in the rest of the paper.
When the Al\textsubscript{2}O\textsubscript{3} seed layer was used,
better film adhesion allowed higher $T_{\textrm{d}}$. These films
were deposited using a linear ramp of $T_{\textrm{d}}$ during deposition,
from $\unit[275]{^{\circ}C}$ to $\unit[400]{^{\circ}C}$ for Cu and
from $\unit[475]{^{\circ}C}$ to $\unit[600]{^{\circ}C}$ for Ni.

In order to evaluate survivability, the films were annealed in a hot-wall
tube furnace before characterization. The annealing conditions were
similar to those used for graphene CVD, but without the hydrocarbon
precursor: duration 10 min to 15 min, temperature of $\unit[1000]{^{\circ}C}$,
flowing 5~\% H\textsubscript{2} in Ar at a total pressure of $\unit[270]{Pa}$
to $\unit[5300]{Pa}$ ($\unit[2]{Torr}$ to $\unit[40]{Torr}$). The
temperature of the furnace was measured using a thermocouple mounted
just outside the quartz tube.

\section{Film Characterization}

\begin{figure}[b]
\centering{}\includegraphics[scale=0.18]{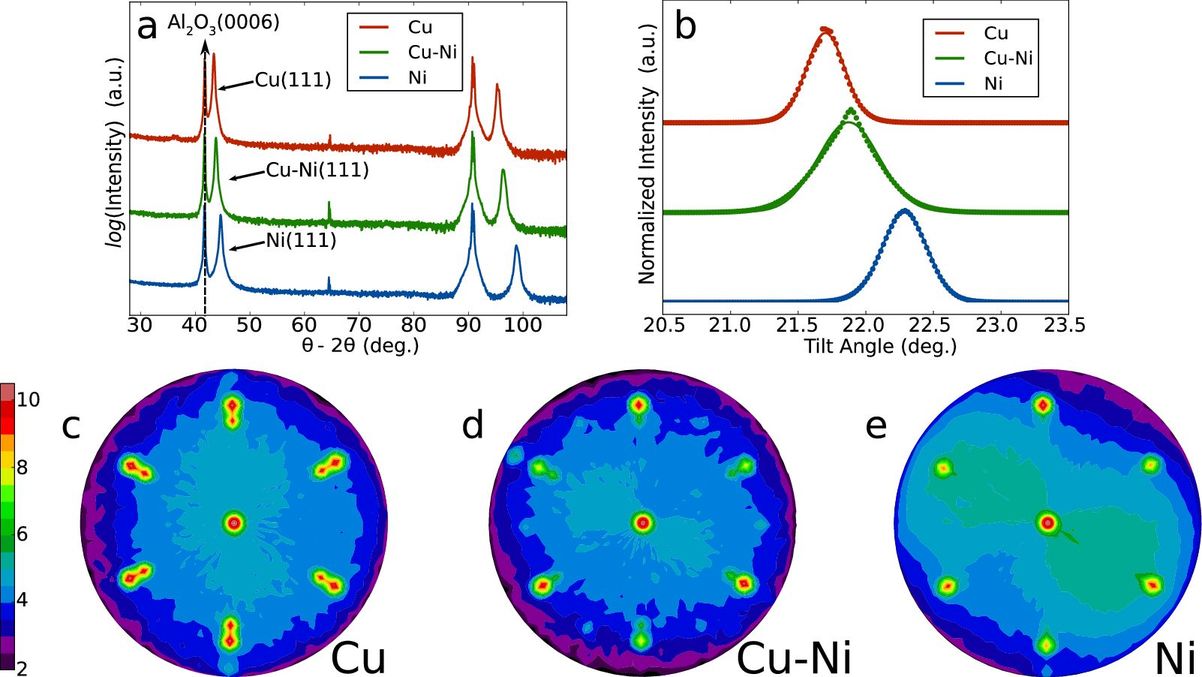}
\caption{Xray diffraction data for $\unit[400]{nm}$ films of Cu, Cu-Ni, and
Ni, deposited without Al\textsubscript{2}O\textsubscript{3} seed
layer, after annealing at $\unit[1000]{^{\circ}C}$ and $\unit[270]{Pa}$
($\unit[2]{Torr}$). (a) $\theta-2\theta$ curves, plotted on a logarithmic
intensity scale, show that all films are exclusively (111) textured.
(b) Rocking curves with Gaussian fits yielding full-width-at-half-maximum
of 0.334\textdegree{} for Cu, 0.533\textdegree{} for Cu-Ni, and 0.393\textdegree{}
for Ni. (c, d, e) (111) pole figures plotted on a logarithmic color
scale. The six innermost spots come from the substrate and the other
spots come from the metal film. The six outer spots of roughly equal
intensity indicate each film contains two families of (111) grains
differing by an in-plane rotation of $60^{\circ}$.}
\label{Xray_data}
\end{figure}

The overall crystalline structure of the films without a seed layer,
as characterized by XRD using Cu $K\alpha$ radiation with a spot
size of several mm\textsuperscript{2}, is shown in Fig. \ref{Xray_data}.
The $\theta-2\theta$ curves in Fig. \ref{Xray_data}a, with intensity
plotted on a logarithmic scale, show a peak at $41.6{}^{\circ}$ from
the Al\textsubscript{2}O\textsubscript{3}(0001) substrate and a
peak corresponding to (111) orientation of the metal film. These peaks
are located at $43.4{}^{\circ}$ for Cu, $43.8{}^{\circ}$ for Cu-Ni,
and $44.6{}^{\circ}$ for Ni. The peaks between $90{}^{\circ}$ and
$100{}^{\circ}$ are higher-order reflections, and the lack of any
other peaks indicates all films are exclusively (111) textured. The
rocking curves in Fig. \ref{Xray_data}b were fit with a gaussian
function, yielding a full-width-at-half-maximum of 0.334\textdegree{}
for Cu, 0.533\textdegree{} for Cu-Ni, and 0.393\textdegree{} for Ni.\textcolor{green}{{}
}The single peak observed for the Cu-Ni alloy film indicates complete
mixing of the two metals during annealing. Its position corresponds
to an alloy composition of Cu\textsubscript{70}Ni\textsubscript{30},
whereas the Cu and Ni film thicknesses predict a composition of Cu\textsubscript{87}Ni\textsubscript{13}.
We attribute this difference to the preferential evaporation of Cu
during annealing. 

Figures \ref{Xray_data}c, d, and e show pole figures measured at
the peak of the (111) reflection and plotted on a logarithmic color
scale. In each pole figure, the six innermost spots come from the
substrate, the other spots come from the metal film, and the rotational
alignment of the two sets of spots indicates the film is epitaxially
aligned with the substrate\textcolor{black}{. The measured epitaxial
relationship for all films can be expressed as $(111)_{\textrm{metal}}$||$(0001)_{\textrm{Al}{}_{2}\textrm{O}_{3}}$
and $[110]_{\textrm{metal}}$||$[10\bar{1}0]_{\textrm{Al}{}_{2}\textrm{O}_{3}}$,
consistent with prior results for several fcc metals on }Al\textsubscript{2}O\textsubscript{3}\textcolor{black}{{}
\cite{Bialas:1994fk}. For a (111) film with a single in-plane orientation,
the pole figure would show only three spots. The six spots with similar
intensity shown in }Fig. \ref{Xray_data}c indicate roughly half the
Cu film is rotated in-plane by $60^{\circ}$ with respect to the rest
of the film. The second set of spots appears with lower intensity
in Figs. \ref{Xray_data}c, d, indicating that only a minority of
the Cu-Ni and Ni films are rotated by $60^{\circ}$. Comparing the
integrated\textcolor{black}{{} intensities for the two sets of spots,
we estimate the fraction of the film }composed of these\textcolor{black}{{}
}$60^{\circ}$ twin domains\textcolor{black}{{} to be 50}~\textcolor{black}{\%
for Cu, 15}~\textcolor{black}{\% for Cu-Ni, and 25}~\textcolor{black}{\%
for Ni. }The Cu-Ni film also shows faint spots in between the six
stronger spots, indicating that a small minority (< 1~\%) of the
film is rotated in-plane by $\pm$30\textdegree{}. Overall, the XRD
characterization shows that all films are (111) textured but have
some degree of in-plane rotational disorder. This is not unexpected
because there are two equivalent orientations for a cubic film deposited
on a hexagonal substrate.

\begin{figure}[b]
\centering{}\includegraphics[scale=0.22]{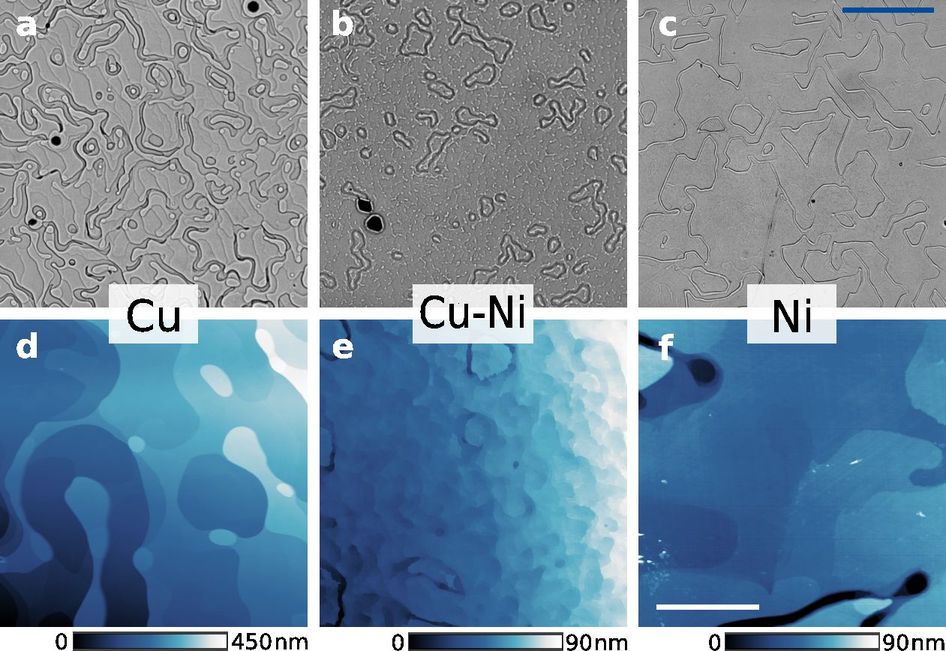}
\caption{Optical and atomic force microscopy images showing the morphology
of the same films as in Fig. \ref{Xray_data}. Top row (a, b, c) shows
optical images ($\unit[50]{\mu m}$ scale bar, shown in (c)) and bottom
row (d, e, f) shows AFM images ($\unit[10]{\mu m}$ scale bar, shown
in (f)) Dark spots in the optical images indicate areas of dewetting.
The Cu surface becomes rough, while the Ni surface remains much smoother.}
\label{Optical_AFM_no_seed}
\end{figure}

The morphology of the films without a seed layer is shown in Fig.
\ref{Optical_AFM_no_seed}. The Cu film shows a rough surface in both
optical and AFM images. The black spots in the optical image are areas
of the film that have dewetted from the substrate, and these spots
increase in size and density for longer annealing times. Both the
roughness and the dewetting indicate the Cu film is only marginally
stable at $\unit[1000]{^{\circ}C}$. Such instability has been noted
previously \cite{Levendorf:2009fk,Ismach:2010bh} and can be attributed
to the rapid evaporation of Cu when heated near its bulk melting temperature
in rough vacuum. This is discussed further below. The Cu-Ni film is
smoother than the Cu film and shows a lower density of dewetting spots.
The Ni film is smoothest of all and has the fewest dewetting spots.\textcolor{black}{{}
}The AFM image of the Ni shows areas of several $\mu\textrm{m}^{2}$
that are atomically smooth (rms roughness $\unit[0.2]{nm}$ to $\unit[0.4]{nm}$
for a $\unit[1]{\mu m}\times\unit[1]{\mu m}$ area). The EBSD characterization
discussed below suggests that the dark lines in the optical images
for Cu-Ni and Ni are boundaries between the $60^{\circ}$ twin domains
revealed in the XRD data. These grain boundaries are wider for the
Cu-Ni film than for the Ni film, and are obscured by roughness for
the Cu film. In the AFM images for Cu-Ni and Ni, the grain boundaries
appear as trenches. These trenches, which do not appear in the as-deposited
films, likely form when grain boundaries move during annealing. These
morphological characterizations show that the Ni film clearly survives
the annealing better than the Cu film, as expected from the higher
melting point of Ni.

\begin{figure}[b]
\begin{centering}
\includegraphics[scale=0.18]{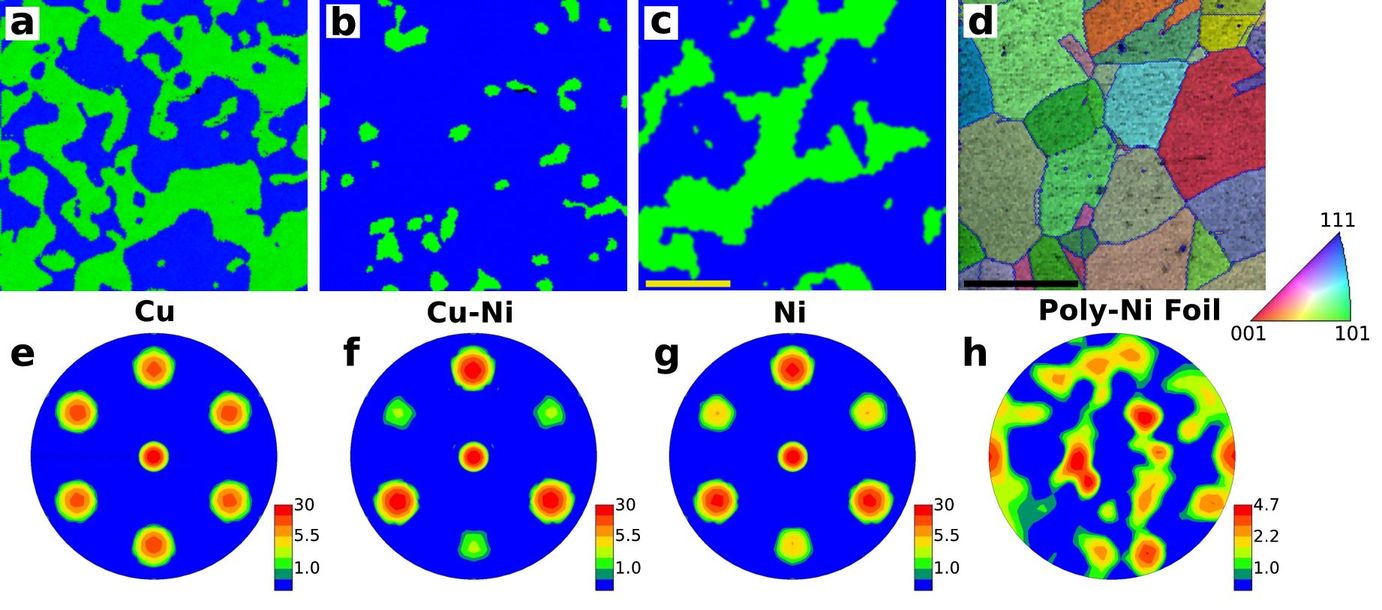}
\par\end{centering}

\centering{}\caption{Electron backscatter diffraction data for the same films as in Fig.
\ref{Xray_data} and for a polycrystalline Ni foil. (a, b, c) Orientation
maps using a color scale that represents in-plane direction only:
blue and green regions differ by $60^{\circ}$. The scale bar shown
in (c) is $\unit[50]{\mu m}$. (e, f, g) (111) pole figures with a
logarithmic color scale. The Cu film shows equal amounts of each orientation,
while the Cu-Ni and Ni films show a majority of one orientation in
the regions shown. (d) Orientation map for a polycrystalline Ni foil
after annealing to $\unit[1000]{^{\circ}C}$. Wedge color scale represents
out-of-plane orientation. Scale bar is $\unit[400]{\mu m}$. (h) (111)
pole figure for the polycrystalline Ni foil.}
\label{EBSD_data}
\end{figure}

The micro-crystalline structure of the films can be characterized
by EBSD, as shown in Fig. \ref{EBSD_data} for films without a seed
layer. Because the films are fully (111) textured, we use a color
scale for in-plane direction only: blue and green regions differ by
a $60^{\circ}$ rotation about the {[}111{]} axis. The upper images
are therefore spatial maps of the $60^{\circ}$ twin domains revealed
in the XRD data. The lower images are (111) pole figures with a logarithmic
color scale. The spatial map for the Cu film shows a broad distribution
of grain sizes, and the pole figure shows equal intensity for both
in-plane orientations. For the Cu-Ni film, both the spatial map and
the pole figure indicate that one orientation dominates in the region
shown here. Comparing the orientation map of Fig. \ref{EBSD_data}b
and the optical image of Fig. \ref{Optical_AFM_no_seed}b, we conclude
that the dark lines enclosing small areas in the optical image are
boundaries between $60^{\circ}$ twin domains. A less pronounced preference
for one in-plane orientation is apparent in the spatial map and pole
figure of the Ni film. For comparison with our (111) films, Figs.
\ref{EBSD_data}d, h show EBSD data for a polycrystalline Ni foil,
also annealed at $\unit[1000]{^{\circ}C}$. In this case we use a
color scale for out-of-plane orientation. The foil shows a broad distribution
of grain sizes and orientations.

\begin{figure}[b]
\begin{centering}
\includegraphics[scale=0.22]{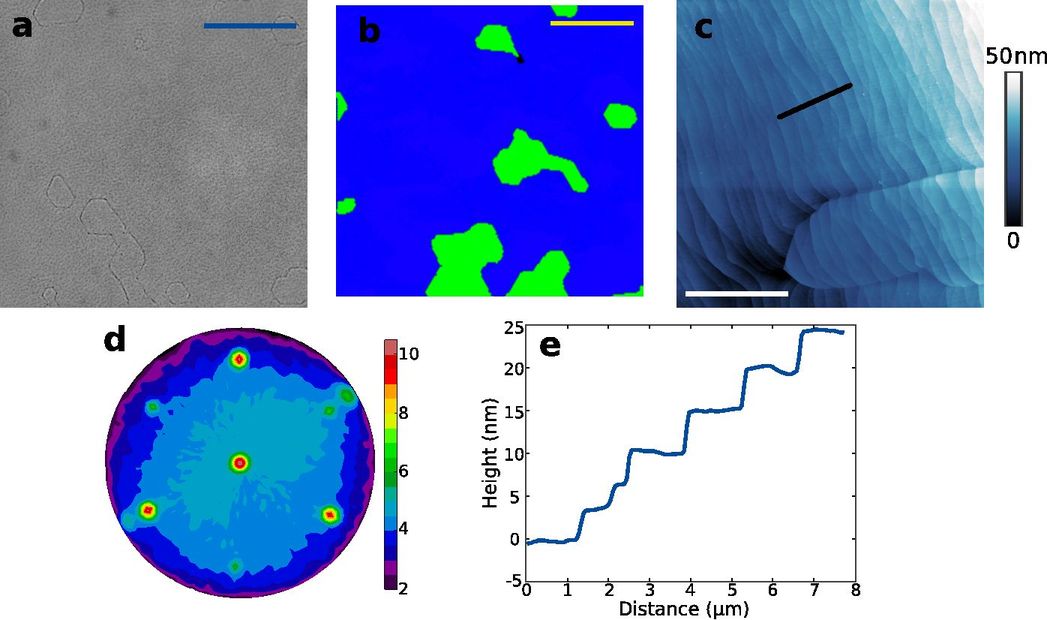}
\par\end{centering}

\centering{}\caption{Crystallinity and morphology of $\unit[400]{nm}$ Ni film deposited
with Al\textsubscript{2}O\textsubscript{3} seed layer, after annealing
at $\unit[1000]{^{\circ}C}$ and $\unit[270]{Pa}$ ($\unit[2]{Torr}$).
Optical image in (a) and EBSD map in (b) ($\unit[50]{\mu m}$ scale
bars) show fewer grains than the corresponding images for Ni without
the seed layer (Figs. \ref{Optical_AFM_no_seed}c and \ref{EBSD_data}c,
respectively). AFM image in (c) shows an ordered surface and shallower
trenches between grains than in Fig. \ref{Optical_AFM_no_seed}f ($\unit[10]{\mu m}$
scale bar). XRD (111) pole figure in (d), plotted on a logarithmic
intensity scale, shows a single in-plane orientation over 97~\% of
the film. Line profile in (e), taken along the black line in (c),
shows steps $\approx\unit[1]{\mu m}$ wide and $\approx\unit[5]{nm}$
high.}
\label{Improved_Ni}
\end{figure}

We now turn to the effect of the Al\textsubscript{2}O\textsubscript{3}
seed layer and other measures that improved the metal films in terms
of epitaxy and/or survivability under graphene CVD conditions. The
seed layer improved the adhesion of all films, as indicated by a reduced
density of dewetting areas after annealing. For the Cu films, the
seed layer had little effect on crystallinity: the XRD pole figure
(not shown) has six spots of similar intensity, indicating roughly
equal amounts of each $60^{\circ}$ twin orientation, as in the data
of Fig. \ref{Xray_data}c. For the Ni films, the seed layer increased
grain size and reduced $60^{\circ}$ twin formation, as shown in Fig.
\ref{Improved_Ni}. The second set of spots in the XRD pole figure
of Fig. \ref{Improved_Ni}d is faint even on a logarithmic scale,
and from their integrated intensity we estimate that only 3~\% of
the film area is composed of $60^{\circ}$ twin domains. The optical
image in Fig. \ref{Improved_Ni}a shows fewer lines due to grain boundaries,
and the lines are less distinct than in Fig. \ref{Optical_AFM_no_seed}c.
Finally, the AFM image in Fig. \ref{Improved_Ni}c shows the detailed
morphology of the improved Ni film. The surface shows a regular series
of atomically flat terraces, also seen in the profile of Fig. \ref{Improved_Ni}e,
that are $\approx\unit[1]{\mu\text{m}}$ wide and separated by steps
$\approx\unit[5]{\text{nm}}$ tall. This profile corresponds to a
slope of 0.3\textdegree{}, consistent with the $\pm0.5^{\circ}$ miscut
specification of the Al\textsubscript{2}O\textsubscript{3} wafer.
The trenches between $60^{\circ}$ twin domains are shallower than
for Ni without the seed layer, which is consistent with less grain
boundary motion during annealing due to the higher deposition temperature.

\begin{figure}[!bh]
\begin{centering}
\includegraphics[scale=0.5]{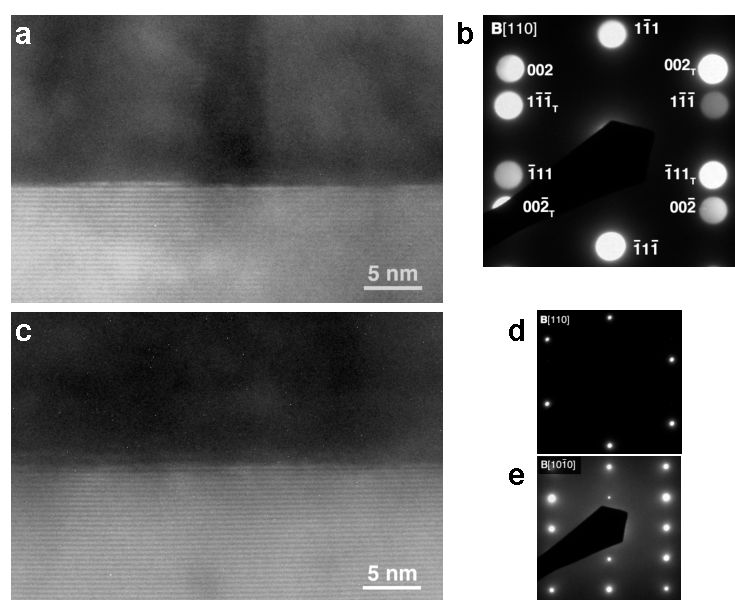}
\par\end{centering}

\centering{}\caption{Cross-sectional transmission electron microscopy of Ni on Al\textsubscript{2}O\textsubscript{3}.
(a, b) Without seed layer. (c, d, e) With seed layer. In (a), a $60^{\circ}$
twin boundary in the Ni is seen at the edge of an extra (0003) Al\textsubscript{2}O\textsubscript{3}
lattice fringe that may indicate a dislocation or step at the surface.
The focused probe diffraction pattern corresponding to the region
of the Ni twin boundary is shown in (b), where the subscript ``T''
labels spots from one of the twin grains. In (c), the lattice fringes
of the film with the seed layer are continuous and smooth, indicating
the surface was not damaged by the resputtering process. Selected
area diffraction patterns are shown for the Ni in (d) and for the
Al\textsubscript{2}O\textsubscript{3} in (e).}
\label{TEM_images}
\end{figure}

To further investigate the effect of the seed layer, TEM was performed
on a cross-section of the Ni/Al\textsubscript{2}O\textsubscript{3}
interface from unannealed samples with and without the seed layer.
The bottom half of each image in Figs. \ref{TEM_images}a,c shows
Al\textsubscript{2}O\textsubscript{3}(0003) lattice fringes, which
are kinematically forbidden but dynamically allowed and are observed
due to double diffraction. Ni lattice fringes are not resolved because
the distance between Ni(111) planes, 0.203 nm, is beyond the 0.28
nm point resolution of the microscope. For the Ni film without the
seed layer, Fig. \ref{TEM_images}a, the last Al\textsubscript{2}O\textsubscript{3}
fringe shows a lateral intensity modulation that could indicate a
fractional-unit-cell island or dislocation near the surface. At the
edge of the extra fringe, there is a vertical grain boundary in the
Ni (darker region near the center of the image). The focused probe
diffraction pattern corresponding to this region, shown in Fig. \ref{TEM_images}b,
indicates both grains share the {[}111{]} out-of-plane and {[}110{]}
in-plane directions. This can only occur for a twin boundary with
a $60^{\circ}$ in-plane rotation about the {[}111{]} direction. The
presence of a Ni grain boundary at the edge of an extra (0003) Al\textsubscript{2}O\textsubscript{3}
lattice fringe was not uncommon, so the extra fringe is likely correlated
with the mechanism for grain boundary formation. Thus the TEM results
independently confirm the EBSD results and provide a microscopic view
of the origin of the $60^{\circ}$ twin domains.

For the Ni film grown with the seed layer, we found far fewer grain
boundaries for an equivalent length of cross-sectional interface.
Fig. \ref{TEM_images}c shows a typical image for this film. The Al\textsubscript{2}O\textsubscript{3}
lattice fringes are continuous and smooth, indicating the reactive
sputtering produced a seed layer with an undamaged surface. Selected
area electron diffraction patterns show the Ni (Fig. \ref{TEM_images}d)
and the Al\textsubscript{2}O\textsubscript{3} (Fig. \ref{TEM_images}e)
are highly ordered.

While the Al\textsubscript{2}O\textsubscript{3} seed layer did not
significantly affect the amount of $60^{\circ}$ twin domains or other
properties of the Cu films, the roughness of these films was strongly
affected by the annealing conditions. The rough surface shown in Fig.
\ref{Optical_AFM_no_seed}a, produced by annealing for 15 min at $\unit[1000]{^{\circ}C}$
in a total pressure of $\unit[270]{Pa}$ ($\unit[2]{Torr}$), offers
a poor substrate for graphene growth. Increasing the total pressure
during annealing reduced the overall roughness of Cu and Cu-Ni films,
as shown in Table \ref{RMS_roughness_table}. We attribute this improvement
to the reduced evaporation rate of Cu at higher pressure. Table \ref{RMS_roughness_table}
also shows that roughness was further decreased by placing a Cu foil
$\approx\unit[1]{\text{mm}}$ above the surface of the Cu film. This
improvement is presumably due to a higher vapor pressure of Cu near
the film, and is likely related to the improvements seen in graphene
CVD when the Cu foil was folded and crimped to make an enclosure \cite{Li:2011nx}.
With both higher total pressure and a foil cover, a Cu(111) film annealed
at $\unit[1065]{^{\circ}C}$ for 15 min showed a roughness only slightly
larger than that of the Ni film in Fig. \ref{Improved_Ni}c. Thus
thin Cu(111) films on Al\textsubscript{2}O\textsubscript{3} can
survive graphene CVD conditions, even those approaching the Cu melting
point, if the total pressure is high enough and there is a nearby
source of Cu vapor.

\noindent 
\begin{table}[h]
\begin{centering}
\textcolor{green}{\small }%
\begin{tabular}{|c|c|c|c|c|c|}
\hline 
\textbf{\textcolor{black}{\small Film}} & \textbf{\textcolor{black}{\small Thickness (nm)}} & \textbf{\textcolor{black}{\small Temp. ($^{\circ}\textrm{C}$)}} & \textbf{\textcolor{black}{\small Press. (Torr)}} & \textbf{\textcolor{black}{\small Foil cover}} & \textbf{\textcolor{black}{\small Roughness (nm)}}\tabularnewline
\hline 
\hline 
\textcolor{black}{\small Cu} & \textcolor{black}{\small 400} & \textcolor{black}{\small 1000} & \textcolor{black}{\small 2} & \textcolor{black}{\small No} & \textcolor{black}{\small 27}\tabularnewline
\hline 
\textcolor{black}{\small Cu} & \textcolor{black}{\small 700} & \textcolor{black}{\small 1000} & \textcolor{black}{\small 20} & \textcolor{black}{\small No} & \textcolor{black}{\small 9.5}\tabularnewline
\hline 
\textcolor{black}{\small Cu} & \textcolor{black}{\small 700} & \textcolor{black}{\small 1000} & \textcolor{black}{\small 40} & \textcolor{black}{\small No} & \textcolor{black}{\small 6.8}\tabularnewline
\hline 
\textcolor{black}{\small Cu} & \textcolor{black}{\small 700} & \textcolor{black}{\small 1000} & \textcolor{black}{\small 20} & \textcolor{black}{\small Yes} & \textcolor{black}{\small 5.8}\tabularnewline
\hline 
\textcolor{black}{\small Cu} & \textcolor{black}{\small 700} & \textcolor{black}{\small 1065} & \textcolor{black}{\small 40} & \textcolor{black}{\small Yes} & \textcolor{black}{\small 5.0}\tabularnewline
\hline 
\textcolor{black}{\small Cu-Ni} & \textcolor{black}{\small 350/50} & \textcolor{black}{\small 1000} & \textcolor{black}{\small 2} & \textcolor{black}{\small No} & \textcolor{black}{\small 18}\tabularnewline
\hline 
\textcolor{black}{\small Cu-Ni} & \textcolor{black}{\small 350/50} & \textcolor{black}{\small 1000} & \textcolor{black}{\small 20} & \textcolor{black}{\small No} & \textcolor{black}{\small 13}\tabularnewline
\hline 
\textcolor{black}{\small Ni} & \textcolor{black}{\small 400} & \textcolor{black}{\small 1000} & \textcolor{black}{\small 2} & \textcolor{black}{\small No} & \textcolor{black}{\small 3.3}\tabularnewline
\hline 
\end{tabular}
\par\end{centering}{\small \par}

\begin{centering}
\caption{RMS roughness of metal films on Al\textsubscript{2}O\textsubscript{3}(0001)
under different annealing conditions. Annealing time was 15 min in
all cases. All roughness values were measured from a $\unit[30]{\mu m}\times\unit[30]{\mu m}$
AFM image.}

\par\end{centering}

\centering{}\label{RMS_roughness_table}
\end{table}

\section{Graphene Growth and Characterization}

Graphene was grown using CH\textsubscript{4} as the precursor gas
in a hot-wall tube furnace with an inner diameter of 80 mm and an
overall length of 1.5 m. The tube was evacuated using a scroll pump
(base pressure of $\unit[2.5]{Pa}$) and all gases were 99.99+~\%
pure. For growth on Ni, the films were annealed at $\unit[1000]{^{\circ}C}$
in a flow of $\unit[27]{\mu mol/s}$ ($\unit[36]{sccm}$) of Ar and
$\unit[27]{\mu mol/s}$ of H\textsubscript{2} for 15 min, cooled
to $\unit[900]{^{\circ}C}$, and then exposed to $\unit[27]{\mu mol/s}$
of H\textsubscript{2} and $\unit[27]{\mu mol/s}$ of CH\textsubscript{4}.
Growth conditions were maintained for 30 min to ensure full C saturation,
and the film was cooled at $\unit[4]{^{\circ}C/min}$ in the same
gas flows used for growth. A total pressure of $\unit[2700]{Pa}$
($\unit[20]{Torr}$) was maintained throughout the process. For growth
on Cu, the films were annealed at $\unit[1035]{^{\circ}C}$ in $\unit[370]{\mu mol/s}$
($\unit[500]{sccm}$) of Ar and $\unit[7.4]{\mu mol/s}$ ($\unit[10]{sccm}$)
of H\textsubscript{2} for 15 min. A flow of $\unit[27]{\mu mol/s}$
($\unit[36]{sccm}$) of CH\textsubscript{4} was then added to the
Ar and H\textsubscript{2}, growth lasted 4 min, and the film was
cooled at approximately $\unit[(50\textrm{ to 100)}]{^{\circ}C/min}$
in the same gas flows used for growth. A total pressure of $\unit[5300]{Pa}$
($\unit[40]{Torr}$) was maintained throughout the process.

\begin{figure}
\begin{centering}
\includegraphics[scale=0.25]{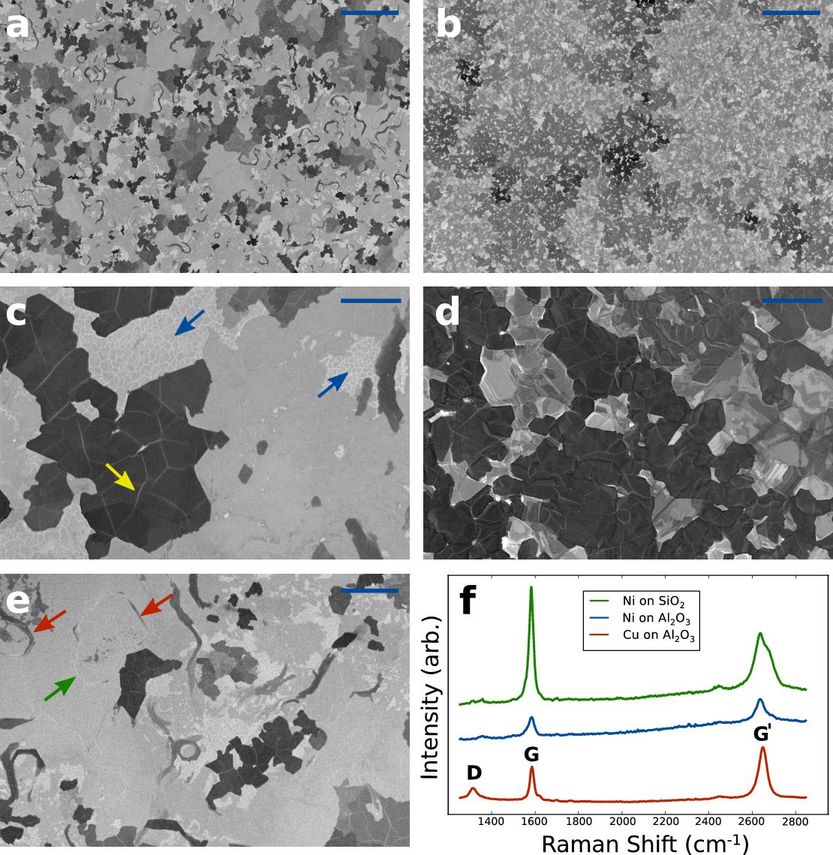}
\par\end{centering}

\centering{}\caption{Scanning electron microscopy (SEM) images of multilayer graphene grown
on Ni films. (a, c, e) $\unit[400]{nm}$ epitaxial Ni(111) on Al\textsubscript{2}O\textsubscript{3}.
(b, d) $\unit[400]{nm}$ polycrystalline Ni on SiO\textsubscript{2}.
Scale bars are $\unit[40]{\mu m}$ for (a) and (b), $\unit[4]{\mu m}$
for (c) and (d), and $\unit[10]{\mu m}$ for (e). (f) Raman spectroscopy
indicates there is at least a monolayer of graphene at all points.
Darker regions correspond to thicker growth.}
\label{SEM_Graphene_on_Ni}
\end{figure}

Scanning electron microscopy (SEM) at low accelerating voltage (5
kV) was used to image graphene on the Ni and Cu surfaces. Figure \ref{SEM_Graphene_on_Ni}
compares growth on $\unit[400]{nm}$ films of epitaxial Ni(111) on
Al\textsubscript{2}O\textsubscript{3} (Fig. \ref{SEM_Graphene_on_Ni}a,c,e)
and polycrystalline Ni on SiO\textsubscript{2} (Fig. \ref{SEM_Graphene_on_Ni}b,d).
The graphene films were grown simultaneously to ensure identical treatment.
We attribute variations in the grey-scale to variations in graphene
thickness, with darker areas presumably corresponding to thicker graphen\textcolor{black}{e.
Although contrast variations from the polycrystalline Ni film itself
could exist due to channeling, we expect this effect to be minimal
because the images were taken with an in-lens detector. The most striking
feature in these images is the different relationship between graphene
thickness changes and Ni grain boundaries for the two films. For the
polyc}rystalline Ni film, most changes in thickness occur at grain
boundaries, leading to mostly faceted edges for regions of a given
thickness (Fig. \ref{SEM_Graphene_on_Ni}d). For the Ni(111) film,
the graphene thickness varies dramatically on a length scale much
smaller than the size of the grains shown in the EBSD data of Fig.
\ref{Improved_Ni}b. Thus while graphene growth on Ni(111) is indeed
more uniform than growth on polycrystalline Ni, it is not nearly as
uniform as one would expect if thickness variations are simply caused
by faster precipitation of C at grain boundaries than within the interior
of grains. This suggests there is an additional source of thickness
variations that remains to be identified. Similar thickness variations
occured for a variety of growth parameters.

The higher resolution SEM images in Fig. \ref{SEM_Graphene_on_Ni}c,e
show detailed features of the thickness variations that are worth
mentioning. In some places, thicker graphene follows the boundary
between a minority grain and the surrounding majority grain (red arrows
in Fig. \ref{SEM_Graphene_on_Ni}e). This has been reported previously
for growth on Ni \cite{Yoshii:2011fk} and is likely due either to
faster precipitation of C or to trapping of C atoms at these boundaries.
In other places, the boundaries of minority grains have no change
in graphene thickness (green arrow in Fig. \ref{SEM_Graphene_on_Ni}e).
In many of the thicker graphene regions, straight, bright lines may
indicate buckling of the graphene from differential contraction during
cooling (yellow arrow in Fig. \ref{SEM_Graphene_on_Ni}c). Some of
the thinnest graphene regions show a network of bright lines that
may indicate a discontinuous mosaic of graphene patches (blue arrows
in Fig. \ref{SEM_Graphene_on_Ni}c). The variety of growth features
on a substrate of such high quality highlights the need for a better
understanding of how graphene grows on Ni.

Raman spectroscopy using an excitation wavelength of 633~nm was done
to verify the existence of graphene and as a rough characterization
of defects. The thickest graphene ($\gtrsim4$ layers) on Ni films
appears darker in an optical microscope. Raman spectra with a spot
size of $\approx\unit[1]{\mu m}\times\unit[3]{\mu m}$ were taken
in between these darker patches to preferentially sample the thinnest
regions. Fig. \ref{SEM_Graphene_on_Ni}f shows typical spectra, which
have the \emph{G} and \emph{G}' peaks characteristic of few-layer
graphene \cite{Malard:2009fk}. No places were found that did not
have these peaks, indicating the films were fully covered by graphene.
The lack of a significant \emph{D} peak in the Raman spectra for graphene
on Ni indicates there are few defects.

The graphene grown on Cu is primarily monolayer, as shown by the Raman
spectrum in Fig. \ref{SEM_Graphene_on_Ni}e, and therefore SEM images
(not shown) reveal little about the growth. The large \emph{D} peak
is likely due to a high density of graphene nucleation sites, resulting
in a large number of graphene grain boundaries. When graphene growth
is interrupted before a complete layer forms, the nucleation sites
can be seen directly in SEM images \cite{Li:2010bh}. Nucleation sites
were separated by roughly $\unit[1]{\mu m}$, much less than the distance
between Cu grain boundaries in Fig. \ref{EBSD_data}a. As with Cu
foils \cite{Li:2010bh}, we expect that optimization of growth conditions
can improve the quality of graphene on Cu(111) films.

\section{Conclusion}

The metal films reported here have many of the properties needed for
wafer-scale growth of graphene by CVD. They can be deposited epitaxially
with a pure (111) texture on commercially available $\alpha-$Al\textsubscript{2}O\textsubscript{3}(0001)
substrates. The films fall short of being single crystals only due
to grain boundaries between (111) regions that differ by an in-plane
rotation of $60^{\circ}$. For Ni, a film that is 97~\% single orientation
and has a smooth, well ordered surface can be achieved by reactive
sputtering of an Al\textsubscript{2}O\textsubscript{3} seed layer
immediately before metal deposition. This process, which probably
helps by removing the OH layer typically found on Al\textsubscript{2}O\textsubscript{3}
after exposure to air, does not require ion bombardment or temperatures
above $\unit[650]{^{\circ}C}$ and thus can be implemented in many
commercial sputtering systems if they are fitted with a basic substrate
heater. Cu films do not show significant improvement from the seed
layer process, but the roughness of these films caused by evaporation
under graphene CVD conditions can be reduced by a combination of higher
total pressure and proximity of a Cu foil. Cu-Ni films deposited as
bilayers become homogeneous alloys when annealed at $\unit[1000]{^{\circ}C}$,
demonstrating the possibility of tuning C solubility through alloy
composition and thereby controlling graphene thickness. The evaporation
of Cu during annealing can have a significant effect on the final
alloy composition.

Further improvements are needed to reach the goal of wafer-scale,
single crystal thin films of Cu, Ni, and Cu-Ni alloys. Optimization
of the reactive sputtering process and the use of a sputtering system
with lower base pressure may result in fewer grain boundaries than
demonstrated here. The improvements found in the Ni films may be transferable
to Cu films through the use of a few nanometers of Ni deposited before
the Cu. Another possible route is the use of MgO substrates. In recent
work, both Cu \cite{Ogawa:2012uq} and Ni \cite{Iwasaki:2011uq} deposited
on MgO(111) showed a (111) texture without apparent in-plane rotational
disorder, and the films did not noticeably degrade under graphene
CVD conditions. These measurements may have been less sensitive to
in-plane rotations than the XRD pole figures presented here, but if
these films do contain a small fraction that is rotated in-plane,
they may also benefit from a reactively sputtered seed layer. Unfortunately,
wafers of crystalline MgO are not widely available.

Independent of progress on film quality, there is clearly room for
improvement in our understanding of graphene CVD on metal substrates,
be they foils, films or ingots. The work presented here implies that
even a single crystal metal film free of grain boundaries may not
be sufficient to achieve the graphene uniformity needed for commercial
applications. The identification of other mechanisms for graphene
nonuniformity remains an important topic of research.

\section{Acknowledgments}

We thank Dustin Hite and David Pappas for advice on metal film growth,
Lawrence Robins and Angela Hight-Walker for assistance with Raman
spectroscopy, and Stephen Russek for helpful discussions about magnetron
sputtering.

\section*{References}
\bibliographystyle{unsrt}
\bibliography{References}

\end{document}